\documentclass[a4paper,fleqn]{cas-sc}

\usepackage[authoryear]{natbib}
\usepackage{upgreek}
\newcommand\micron{$\upmu$m }
\newcommand\CH{CH$_4$}
\newcommand\chf{\textrm{CH}_4}
\newcommand\N{N$_2$}
\newcommand\nf{\textrm{N}_2}
\newcommand\HO{H$_2$O}

\newcommand\TT{Titan Tholin}

\begin{document}
\shorttitle{Pluto’s Surface Composition and Temperature}
\shortauthors{A.E. Drozdov and N.V. Emelyanov}

\title [mode = title]{New surface composition and temperature maps of Pluto from New Horizons LEISA data}                 

\author[1]{A.E. Drozdov}[orcid=0009-0005-5109-7863]
\cormark[1]
\ead{drozdofflekha1997@yandex.ru}

\author[1]{N.V. Emelyanov}[orcid=0000-0002-5974-5125]

\affiliation[1]{organization={M.V. Lomonosov Moscow State University/Sternberg astronomical institute},
                addressline={Universitetskij prospect 13}, 
                city={Moscow},
                postcode={119234}, 
                country={Russia}}

\cortext[cor1]{Corresponding author}

\nonumnote{Accepted for publication in Icarus. DOI: \href{https://doi.org/10.1016/j.icarus.2026.117031}{10.1016/j.icarus.2026.117031}. \copyright 2026. This manuscript version is made available under the CC BY-NC-ND license.}

\begin{abstract}
New Horizons RALPH/LEISA near-infrared spectra allow for regional mapping of Pluto's surface ices and their physical state; however, scan-to-scan artifacts and variable spatial resolution complicate quantitative interpretation. We extend previous LEISA compositional studies \citep{Protopapa2017, Schmitt2016, Emran2023} by combining five close-approach observations into co-registered equirectangular maps at 7 km/pixel and by jointly retrieving surface temperature, ice abundances, and grain sizes using a Hapke-based mixing model.

We mitigate bad pixels and edge overexposure linked to flat-field uncertainties and correct for residual scan-to-scan spectral discrepancies using per-observation scale and offset terms. The resulting maps provide distributions of \CH{}-rich ice, \N{}-rich ice, \HO{} ice, and \TT{}, alongside a corresponding temperature map.
\end{abstract}


\maketitle

\section{Introduction}
New Horizons revealed Pluto as an active, volatile-rich world with strong regional contrasts in geology, albedo and ice composition. Spatially resolved near-infrared spectroscopy is a key tool for connecting these geologic units to the distribution and physical state of volatile ices (\N{}, \CH{}, and CO) and non-volatile components such as \HO{} ice and complex organics (tholins). These surface materials control energy balance and sublimation/condensation cycles and therefore underpin interpretations of surface-atmosphere exchange and volatile transport across Pluto.

The RALPH/LEISA imaging spectrometer acquired spectra in the 1.25--2.5~\micron range (with a higher-resolution subrange near 2.1--2.25~\micron) during the 2015 flyby \citep{Reuter2007}. LEISA data have enabled global compositional mapping and quantitative spectral modelling, including pixel-by-pixel Hapke modelling \citep{Protopapa2017} and alternative approaches based on band analysis, dimensionality-reduction methods \citep{Schmitt2016} and unsupervised learning \citep{Emran2023}. In particular, \citet{Protopapa2017} demonstrated that a compact set of endmembers (\CH{}-rich ice, \N{}-rich ice, \HO{} ice and \TT{}) can reproduce the main spectral variability and yield physically interpretable global maps.

However, two practical limitations motivate the present work. First, individual LEISA scans differ in viewing geometry, signal level and spatial resolution, and combining them into a single map product requires careful co-registration and artifact mitigation. Second, temperature-dependent shifts and broadening of \N{}, \CH{} and \HO{} absorption bands contain information on ice temperature and physical state, but this information is not always carried through into the same pixel-by-pixel compositional framework.

The goal of this work is therefore to produce updated composition maps at 7 km/pixel by combining five close-approach LEISA observations, and to retrieve a surface temperature parameter simultaneously with ice abundances and grain sizes using the Hapke-based approach of \citet{Protopapa2017}. In doing so, we describe a preprocessing workflow that detects and corrects several observation artifacts in the calibrated FITS cubes, and quantify residual scan-to-scan spectral level mismatches via per-observation scale and offset terms.

\section{Methodology}
\subsection{Data set and workflow overview}
To reduce the amount of instrument-specific knowledge required to follow the analysis, we summarize here the inputs and processing steps used throughout the paper.

\textbf{Input data:} calibrated New Horizons RALPH/LEISA FITS cubes (Table~\ref{tbl2}) containing radiometric measurements ($erg/s/cm^2/\mathring{A}/str$).

\textbf{Preprocessing of each LEISA cube:}

\begin{enumerate}[1.]
\item detection of bad pixels using the provided mask and a Kalman-filter-based smooth estimate, followed by interpolation;
\item subtraction of an ``offset field'' estimated by averaging frames without Pluto in the field of view;
\end{enumerate}

\textbf{Multi-cube correction:}

\begin{enumerate}[3.]
\item empirical correction of edge overexposure via a revised flat-field derived by minimizing local contrast consistently across multiple cubes.
\end{enumerate}

\textbf{Map generation:} conversion of each cube into co-registered equirectangular maps at 7 km/pixel using ISIS USGS, including per-pixel geometry (incidence, emission, phase angles).

\textbf{Spectral modeling and inversion:} pixel-by-pixel Hapke-based mixing model with free parameters $F_{1-4}$, $D_{1-4}$, $f_{\chf:\nf}$ and $T$, solved by Levenberg--Marquardt (with both least squares and $\chi^2$ objective functions tested). To account for scan-to-scan spectral level mismatches, each observation is allowed an additional scale and offset $(S_j,O_j)$.

\subsection{Spectral modeling}
We modeled the bidirectional reflectance for each material component using Hapke's law \citep{Hapke2012}:

\begin{equation}
r(i,e,\phi)=\frac{w}{4\pi}\frac{\mu'_0}{\mu'_0+\mu'}\left\{\left[1+B(\phi,B_0,h)\right]P(\phi,g)+H(\mu',w)H(\mu'_0,w)-1\right\}S(i,e,\overline{\theta}).
\end{equation}

Here, $i, e$, and $\phi$ are the incidence, emission, and phase angles, respectively; $w$ is the single scattering albedo; and $\overline{\theta}$ represents the average slope of the macroscopic surface roughness. The opposition effect is described by:

\begin{equation}
B(\phi,B_0,h)=\frac{B_0}{1+\frac{tg(\phi/2)}{h}}
\end{equation}
where $B_0$ is the amplitude and $h$ is the angular half-width of the opposition effect. The single particle phase function is approximated using the Henyey-Greenstein relation:

\begin{equation}
P(\phi,g)=\frac{1-g^2}{\left(1+2g\cos\phi+g^2\right)^{3/2}},
\end{equation}
where $g$ is the asymmetry parameter. For the multiple scattering function $H(\mu',w)$, we utilized the Ambartsumian-Chandrasekhar approximation (Eq. 8.53 in \citep{Hapke2012}):

\begin{equation}
H(\mu',w)=\frac{1+2\mu'}{1+2\mu'\sqrt{1-w}}.
\end{equation}
Following Hapke's terminology, we model the radiance factor (denoted by RADF or $\frac{I}{F}$). The final bulk radiance factor is obtained via a linear combination of the individual material components, weighted by their fractional area:

\begin{equation}
\frac{I}{F}=\pi\sum_iF_ir_i, where \sum_iF_i=1
\end{equation}
The four macroscopic Hapke parameters ($g, h, B_0, \overline{\theta}$) were assumed to be wavelength-independent and uniform across all materials. We adopted the following values: $g = -0.3$, $h=0.5$, $B_0=1.0$, and $\overline{\theta}=10^\circ$ \citep{Olkin2007, Buie2010}. As the simultaneous retrieval of all Hapke parameters is a mathematically ill-posed problem—where different parameter combinations often yield identical reflectance values—we fixed these parameters to ensure a unique solution.
To model the wavelength dependence of the single scattering albedo, $w$, we employed the equivalent slab model \citep{Hapke2012}. This model relates the albedo to the effective particle diameter, $D$, and the optical constants—specifically, the real ($n$) and imaginary ($k$) parts of the refractive index:

\begin{equation}
w(n,k,D)=S_e+\left(1-S_e\right)\left(1-S_i\right)\frac{e^{-\alpha D}}{1-S_ie^{-\alpha D}},
\end{equation}
where: 

\begin{equation}
\left.
\begin{aligned}
& S_e=0.0587+0.8543\,R_0+0.087\,R_0^2,\\
& S_i=1 - \left(0.9413-0.8543\,R_0-0.087\,R_0^2\right)/n^2,\\
& R_0=\frac{(n-1)^2+k^2}{(n+1)^2+k^2},
\end{aligned}
\right.
\end{equation}
and $\alpha = 4\pi k(T)/\lambda$ is the absorption coefficient. The fractional areal coverage, $F_i$, and the effective particle diameter, $D_i$, serve as free parameters in the model. The refractive indices $n(T)$ and $k(T)$ are temperature-dependent; therefore, the temperature, $T$, is also treated as a free parameter. The wavelength-dependent optical constants for each material at various temperatures we used from laboratory measurements (Table \ref{tbl1}).

Following \citet{Protopapa2017}, we modeled the spectral contributions of four materials: \CH{}-rich ice, \N{}-rich ice, \HO{}-ice, and \TT{}. Since complete optical constants for \N{}-rich ice containing \CH{} impurities at low temperatures are unavailable, we approximated them using a weighted sum of the spectra for pure \N{} ice and for \CH{} dissolved in \N{}, following the method of \citet{Doute1999}:

\begin{equation}
n_{\nf:\chf}=n_{\nf} \left(1-f_{\chf:\nf}\right)+n_{\chf\,in\,\nf}\,\,f_{\chf:\nf}
\end{equation}
where $f_{\chf:\nf}$ represents the concentration of \CH{} in \N{}. The imaginary part, $k_{\nf:\chf}$, is calculated analogously.

Our calculations initially relied on \citet{ProkhvatilovPhase}, which suggests a solubility limit of \CH{} in \N{} of $\le 5\%$ at relevant temperatures. A recent study by \citet{RAPOSA2026} has since identified a previously unknown sub-solidus phase boundary near 53 K, indicating that the low-temperature phase behavior of this binary system is more complex than previously understood. While this discovery may eventually lead to refinements of the exact solubility limits at 40 K, quantifying this shift requires further thermodynamic modeling. Therefore, we maintain the $\le 5\%$ limit as our baseline assumption for the current calculations. Consequently, we treated the concentration $f_{\chf:\nf}$ as a free parameter. The low-temperature spectra for \N{} ice and dissolved \CH{} are listed in Table \ref{tbl1}.

\begin{table*}
\caption{Optical constants used in work}
\resizebox{\textwidth}{!}{%
\begin{tabular}{l l l l l p{3.2cm}}
\hline
Material   & Source                  & SSHADE Source\footnotemark[1] & Temperature & Wavelength range & Notes\\
           &                         &                               & (K)         & ($\upmu$m)       &\\
\hline
\\
\TT{}        & \citet{Khare1984}       & -                                                                                         & 39        & 0.02 - 920.0 &\\
\\
\CH{}-ice    & \citet{Grundy2002CH4}   & \href{https://www.sshade.eu/data/SPECTRUM_BS_20130114_003}{CH4 phase-I,-II crystal}       & 20 - 93   & 0.7 - 5.0    & \\
\\
\N{}-ice     & \citet{Grundy1993N2}    & \href{https://www.sshade.eu/data/SPECTRUM_BS_20120925_003}{$\alpha$,$\beta$-N2 crystal}   & 33 - 43 & 2.062 - 4.762& We set $n=1.23$ and $k=0$, below 2.1~\micron.\\
           & \citet{Tryka1995N2Temp} & -                                                                                         & 35.6 - 58.3 & 1 - 5    & \\
\\
\N{}:\CH{}-ice & \citet{Quirico1999N2CH4}& \href{https://www.sshade.eu/data/SPECTRUM_BS_20130103_003}{CH4 in solid solution $\alpha$,$\beta$-N2} & 35 - 43 & 1 - 5        & $\alpha$- and $\beta$-\N{}:\CH{}~solid solution with \CH{}~concentrations lower than 2\%. The absorption coefficient of diluted \CH{}~is normalized to a concentration of 1.\\
\\
\HO{}-ice    & \citet{Grundy1998H20}   & \href{https://www.sshade.eu/data/SPECTRUM_BS_20120924_012}{H2O Ih crystal}                & 20 - 293  & 0.96 - 2.74  &\\
\hline
\end{tabular}
}
\footnotemark[1]{Optical constants are available at \url{https://www.sshade.eu/}.}
\label{tbl1}
\end{table*}

\subsection{Processing observations}

All LEISA infrared spectral images used in this study are publicly available and were retrieved from the NASA Planetary Data System: \url{https://pdssbn.astro.umd.edu/holdings/nh-p-leisa-3-pluto-v3.0/}. The data are radiometrically calibrated, providing spectral radiance in units of $erg/s/cm^2/\mathring{A}/str$.

The LEISA instrument utilizes a linear variable filter (LVF). Each FITS file consists of an image cube generated by scanning over a specific exposure interval. In the FITS coordinate system, the $x$-axis corresponds to the spatial cross-track dimension, while the $y$-axis represents both the spatial along-track dimension and the spectral dimension. The instrument covers two wavelength ranges: 1.25--2.5 \micron with a spectral resolving power of $R = \lambda/\Delta\lambda = 240$, and a higher-resolution segment from 2.1--2.25 \micron with $R = 560$. For this analysis, we selected the five LEISA observations taken at the closest distances to Pluto (Table \ref{tbl2}):

\begin{table}
\caption{LEISA observations used for calculation in this work.}
%\resizebox{\textwidth}{!}{%
\centering
\footnotesize
\begin{tabular}{l l l l p{3.2cm}}
\hline
File name                   & S/C start time, UTC  & Phase       & Distance & Spatial scale\footnotemark[2] \\
                            &                      & ($^{\circ}$) & (km)     & (km/pix)  \\
\hline
\\
lsb\_0299169338\_0x53c\_sci & 2015-07-14  08:43:40 & 20.3        & 149712   & 9.1   \\
\\
lsb\_0299170159\_0x53c\_sci & 2015-07-14  08:57:21 & 20.7        & 139433   & 8.5   \\
\\
lsb\_0299172014\_0x53c\_sci & 2015-07-14  09:28:16 & 22.0        & 112744   & 6.8   \\
\\
lsb\_0299172889\_0x53c\_sci & 2015-07-14  09:42:51 & 22.8        & 100298   & 6.1   \\
\\
lsb\_0299176809\_0x53c\_sci & 2015-07-14  10:48:11 & 31.8        & 45226    & 2.6   \\
\hline
\\
\footnotemark[3]lsb\_0299105209\_0x54b\_sci & 2015-07-13  14:54:51 & 15.8        & 1032787  & 64.0 \\
\\
\footnotemark[3]lsb\_0299127869\_0x53c\_sci & 2015-07-13  21:12:31 & 16.2        & 718988   & 44.4  \\
\\
\footnotemark[3]lsb\_0299144829\_0x54b\_sci & 2015-07-14  01:55:11 & 16.7        & 485135   & 30.0  \\
\hline
\end{tabular}

\footnotemark[2]{Calculated with ISIS USGS utility (\url{https://isis.astrogeology.usgs.gov/})} \\
\footnotemark[3]{Files was additionally used only for calibration procedure}
\label{tbl2}
\end{table}

For the selected FITS files, we performed preprocessing associated with the detection of bad pixels and their interpolation by the neighbor's pixels values. In addition to the standard bad pixel mask from original images, we implemented a Kalman filter to identify transient artifacts. The filter generated a temporally smooth estimate for each pixel within the FITS cube. Individual measurements deviating from this smoothed estimate by more than $3\sigma$ were flagged as bad pixels. Notably, the Kalman-smoothed data were utilized solely for outlier detection; the original radiometric values were preserved for subsequent analysis. Then for all bad pixels we interpolated their values by neighbor's pixels.

\begin{figure*}
	\centering
	\includegraphics[width=1.0\linewidth]{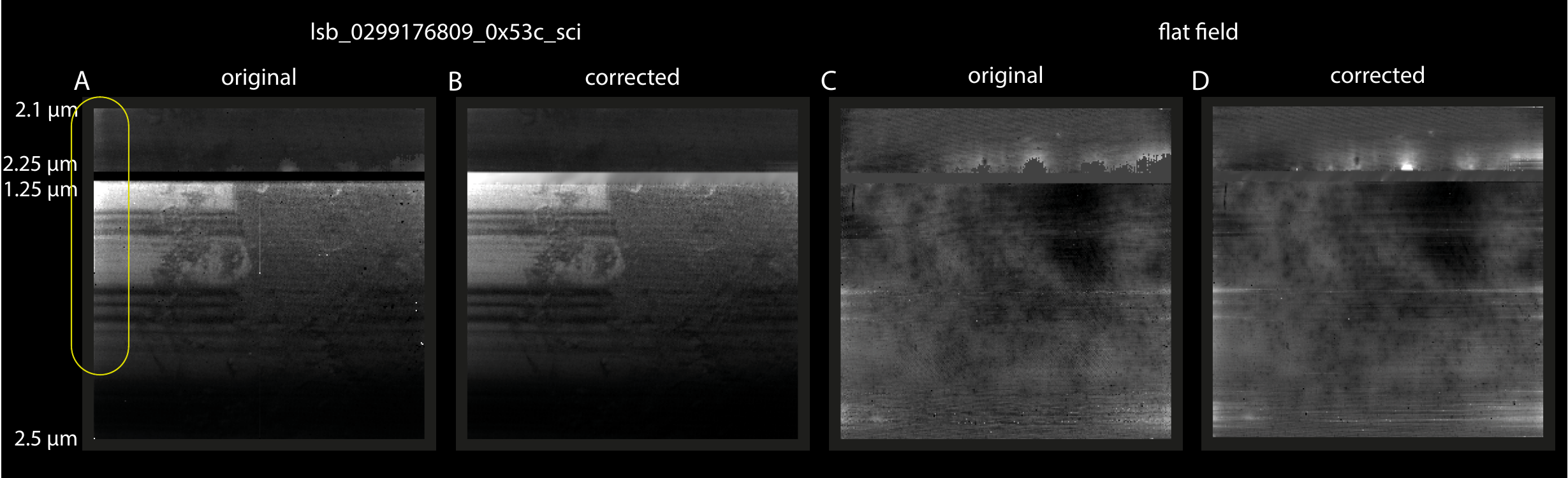}
	\caption{Example of preprocessing and flat-field correction for the closest LEISA scan (\texttt{lsb\_0299176809\_0x53c\_sci}). Panel A shows the original calibrated frame (reflectance level still affected by edge overexposure), while panel B shows the same frame after bad-pixel interpolation, offset field subtraction, and the empirical flat-field correction described in Section~2.3. Panels C and D show the original flat field $f(x,y)$ and the corrected flat field $f^*(x,y)=f(x,y)e^{\beta(x,y)}$, respectively; the strongest adjustments occur near the left detector edge, where the overexposure artifact is most pronounced.}
	\label{FIG:flat}
\end{figure*}

During the preliminary analysis of the FITS data, intensity artifacts were identified along the edges of the images, particularly on the left side (Fig. \ref{FIG:flat}A). This artificial brightening significantly distorts photometric measurements and is present across all files; a similar, though less pronounced, effect is observed on the right edge.

To mitigate this, an offset field—calculated by averaging frames where Pluto was absent—was subtracted from each image. Analysis of this offset field relative to the measurement noise (empirical NESR, estimated as the standard deviation of the background pixels) revealed a distinct spatial gradient across the detector. In the central region, the magnitude of the offset was generally comparable to or slightly lower than the noise level (signal/noise ratio vary from 0 to 2). However, towards the left and bottom edges of the array, the systematic offset systematically exceeded the random measurement noise by a factor of 1 to 5. Consequently, subtracting this two-dimensional offset field was crucial to remove spatially dependent systematic biases (likely associated with instrumental background or readout electronics) that would otherwise contaminate the target's signal. However, this subtraction did not fully eliminate the edge artifacts, suggesting the issue stems from inaccuracies in the flat-field correction (pixel sensitivity map) rather than a simple additive background. While the original flat field corrects for pixel sensitivity variations across most of the detector, it appears to be imprecise at the extreme edges. Consequently, we developed the following algorithm to refine the flat-field correction.

The standard radiometric calibration is derived from the raw measurements using the expression defined in the \texttt{EXTENSION\_GAIN\_OFFSET} section of the associated label (\texttt{.lbl}) file:

\begin{equation}
I_i(x,y,t)=G(x,y)\frac{D_i(x,y,t)-B_{i}(x,y)}{f(x,y)\,\Delta t_i}
\end{equation}
Here, $D_i(x,y,t)$ represents the raw data numbers (DN) for the $i$-th FITS file; $B_{i}(x,y)$ is the subtracted bias (offset); $G(x,y)$ is the gain factor for converting to radiometric values; $\Delta t_i$ is the exposure time; and $f(x,y)$ denotes the flat-field response. The coordinates $(x,y)$ correspond to the spatial dimensions of the FITS image (cross-track and along-track/spectral, respectively), while $t$ represents the frame index.

To linearize the problem, we computed the natural logarithm of the time-averaged images for each file:

\begin{equation}
L_i(x,y)=ln\left(\frac{\overline{D_i(x,y)}}{f(x,y)\,\Delta t_i}\right),
\end{equation}
\begin{equation}
\overline{D_i(x,y)}=\sum_t \left(D_i(x,y,t)-B_{i}(x,y)\right)/N_t
\end{equation}
This logarithmic transformation converts the problem of finding a multiplicative flat-field correction factor into the simpler task of determining an additive shift term. 

Next, we employed the Levenberg-Marquardt algorithm to minimize the following objective function:

\begin{equation}
\label{eq:minfunc}
F(\beta)=\prod_i\left[\sum_{\substack{p=-10\\ p\neq0}}^{10}\left(L_i(x+p,y)-L_i(x,y)+|\beta|\right)^2\right]
\end{equation}
The Levenberg-Marquardt algorithm was selected for its robust convergence and ease of implementation. The minimization is performed independently for each pixel $(x,y)$ across all files, solving for a single scalar parameter $\beta$—the optimal logarithmic offset. After convergence, the resulting $\beta(x,y)$ is used to compute the refined flat field $f^*(x,y)$.

\begin{equation}
ln\left(\frac{\overline{D_i}}{f^*\,\Delta t_i}\right) = ln\left(\frac{\overline{D_i}}{f\,\Delta t_i}\right)-\beta,
\end{equation}
\begin{equation}
f^*(x,y)=f(x,y)\,e^{\beta(x,y)}
\end{equation}

The objective function (Eq. \ref{eq:minfunc}) was designed to minimize local image variance—specifically, the sum of squared differences between a pixel and its neighbors within a small window along the spatial x-axis. By minimizing the product of these contrast terms across all FITS files, the algorithm acts as a coincidence filter: a pixel's value is significantly adjusted only if it deviates from its neighbors consistently across all observations. 

This approach relies on the assumption that the flat-field error pattern ($\beta(x,y)$) is static, whereas real scene features (such as the limb of Pluto) shift position between exposures. Consequently, if a high-contrast feature appears in one file but shifts in others, the product term remains small, and the pixel value is preserved. To ensure the effective separation of static errors from dynamic features, the dataset must include observations with varying target positions. Therefore, we augmented the initial set of five files with three additional observations (listed in Table \ref{tbl2}).

\begin{figure*}
	\centering
	\includegraphics[width=1.0\linewidth]{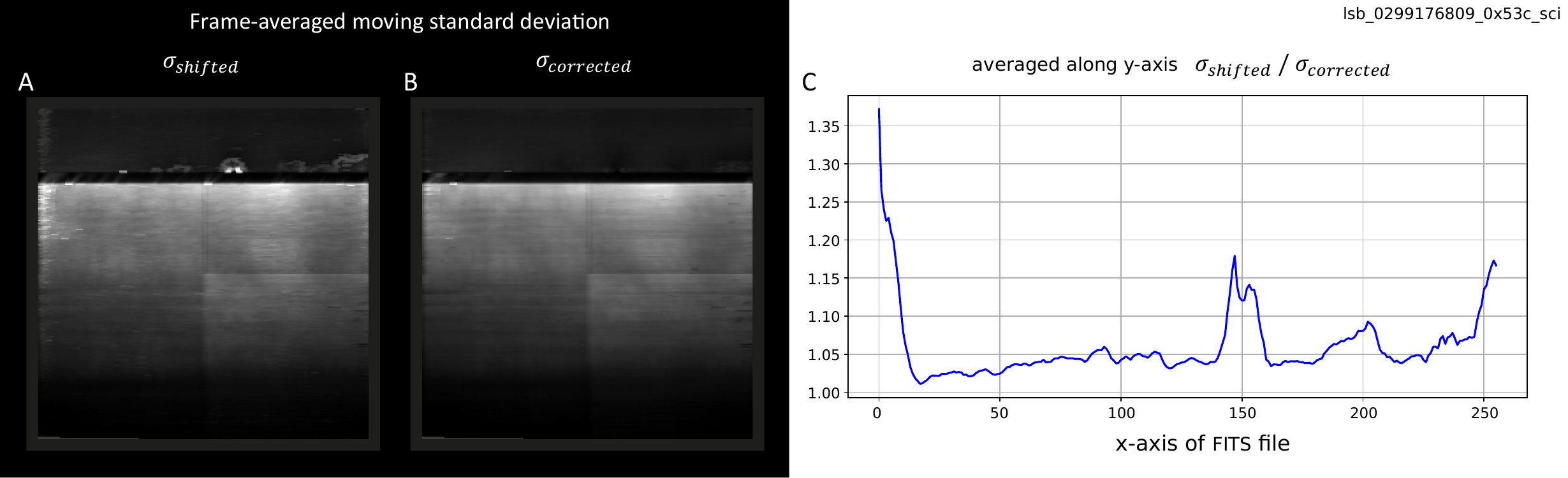}
	\caption{Frame-averaged moving standard deviation along the x-axis using a sliding window of 7 pixels for the LEISA scan (\texttt{lsb\_0299176809\_0x53c\_sci}). Panel A shows frame-averaged moving standard deviation ($\sigma_{shifted}$) for the scan after bad-pixel correction and offset field subtraction, while panel B shows the same quantity ($\sigma_{corrected}$) after the flat-field correction. Panel C shows the ratio $\frac{\sigma_{shifted}}{\sigma_{corrected}}$ averaged along the FITS y-axis, indicating that the flat-field correction strongly reduces large values of $\sigma$ in overexposed regions (e.g., by 5--35\% near the left edge at $x<20$, and by 5--15\% near $x\approx150$ and at the right edge), while slightly reducing $\sigma$ everywhere by $\sim$2.5\% on average due to smoothing of the flat field.}
	\label{FIG:flat2}
\end{figure*}

We acknowledge that this method has limitations. Specifically, solving for the parameter $\beta$ for each pixel independently effectively smooths the flat field, which may introduce artificial softening to the data. To quantify this effect, we calculated the moving standard deviation within a 7-pixel sliding window along the x-axis. We compared the deviation before applying the new flat field ($\sigma_{shifted}$, Fig. \ref{FIG:flat2}A)—calculated after bad-pixel correction and offset-field subtraction—against the deviation after correction ($\sigma_{corrected}$, Fig. \ref{FIG:flat2}B). Figure \ref{FIG:flat2}C displays the ratio $\sigma_{shifted}/\sigma_{corrected}$ averaged along the y-axis. The results indicate a baseline reduction in standard deviation of approximately 2.5\%, suggesting a minor loss of high-frequency detail.

The smoothing window was applied along the x-axis (spatial cross-track) because the y-axis corresponds to different wavelengths in the LEISA scanning geometry. It should be noted that due to the "spectral smile" effect—where wavelength bands curve across the detector—a single row (constant $y$) is not strictly monochromatic. However, despite these caveats, the algorithm is computationally efficient and significantly mitigates flat-field distortions. As shown in Fig. \ref{FIG:flat2}C, the variance is reduced by up to 35\% in high-noise or overexposed regions, effectively suppressing artifacts while minimizing signal loss in nominal regions.

In summary, the data preprocessing pipeline consisted of the following steps:

\noindent \textit{For each of the 5 individual measurement files:}
\begin{enumerate}[1.]
    \item Identification of bad pixels using the instrument mask and a Kalman filter-based smoothing estimator.
    \item Spatial interpolation of bad pixels using neighboring values.
    \item Subtraction of the background ``offset field'' derived by averaging frames where Pluto was outside the field of view.
\end{enumerate}

\noindent \textit{For the combined dataset:}
\begin{enumerate}[4.]
    \item Computation and application of refined flat-field corrections to mitigate detector non-uniformities and high-signal artifacts.
\end{enumerate}

Figure \ref{FIG:flat} illustrates the results of this pipeline for the FITS file \texttt{lsb\_0299176809\_0x53c\_sci} (Panels A and B), and compares the original versus the refined flat field (Panels C and D).

Finally, to derive the radiance factor (or reflectance, $I/F$), the calibrated spectral radiance was normalized by the solar irradiance spectrum. We utilized the solar spectrum model from \citet{SolarIss} (available at \url{https://cdsarc.cds.unistra.fr/ftp/J/A+A/611/A1/}), scaled to the heliocentric distance of Pluto at the time of observation.

\subsection{Map projection}
\begin{figure*}
	\centering
	\includegraphics[width=1.0\linewidth]{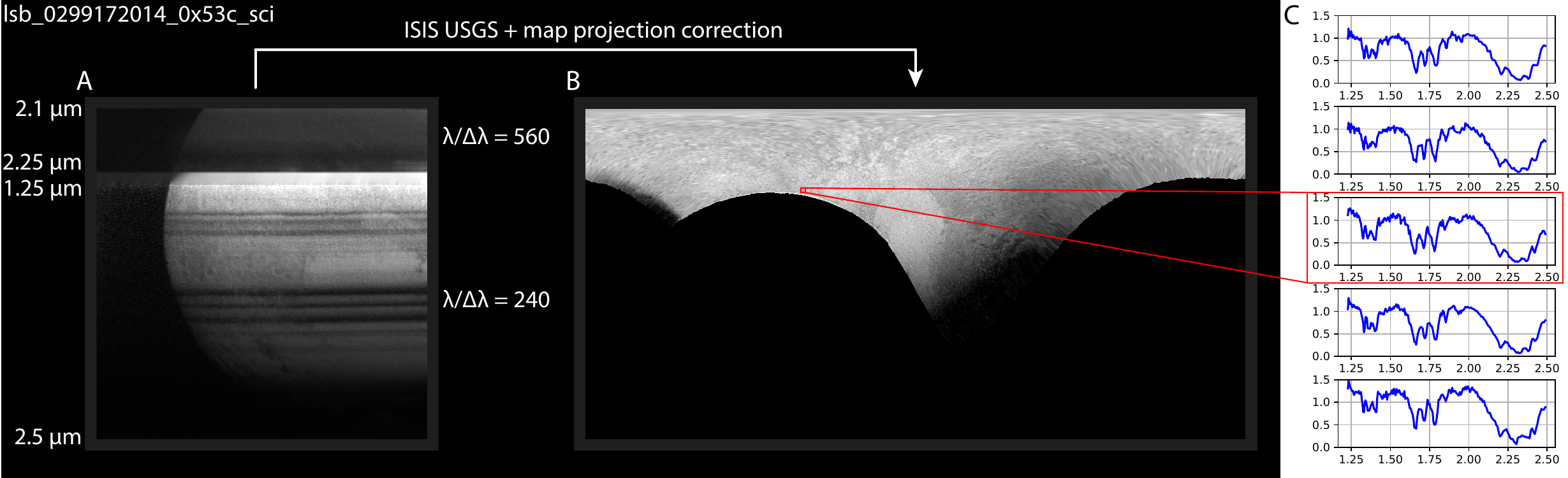}
	\caption{Example of map processing from LEISA scan. Panel A shows FITS file (\texttt{lsb\_0299172014\_0x53c\_sci}) after preprocessions described in section~2.3, panel B shows equirectangular map projection, obtained from FITS file by applying ISIS USGS utility and map corrections, described in section~2.4. Panel C shows a set of measurements for one pixel in projection map. From 5 FITS files (Table \ref{tbl2}) we have 5 maps and so 5 spectra for one pixel. However, due to the fact that the maps cover different parts of Pluto, the count of spectra for different pixels vary from 0 to 5.}
	\label{FIG:map_proj}
\end{figure*}

Following the corrections, we generated equirectangular projection maps for each wavelength at a spatial resolution of 7 km/pixel (Fig. \ref{FIG:map_proj}B). This processing was performed using the USGS Integrated Software for Imagers and Spectrometers (ISIS) (\url{https://isis.astrogeology.usgs.gov/}), which natively supports New Horizons LEISA data.

Leveraging the instrument kernels and spacecraft trajectory data (SPICE), ISIS computes not only the geographic projections but also the per-pixel photometric geometry—specifically the angles of incidence ($i$), emission ($e$), and phase ($\phi$)—required for spectral modelling. 

Although ISIS provides initial georeferencing, the resulting five projection maps exhibited slight spatial misalignments; identical surface features (e.g., craters) appeared at varying coordinates across the different observations. To correct for these offsets and coregister the maps, we applied the dense optical flow algorithm of \citet{Farneback2003}. This method computes the displacement field between images and is available via the \texttt{calcOpticalFlowFarneback} function in the Python OpenCV library (\texttt{cv2}).

A global high-resolution LORRI mosaic of Pluto (file \texttt{nh\_pluto\_mosaic.img}; \citet{Stern2025}) served as the primary geometric reference for co-registering the LEISA observations. However, due to limited surface contrast in the LORRI mosaic at high northern latitudes, this template proved inadequate for precise alignment in the polar regions. Consequently, we adopted the LEISA observation \texttt{lsb\_0299172014\_0x53c\_sci} as the reference anchor for the north pole, as it was acquired near closest approach and offers superior polar coverage. Specifically, the alignment was performed using the 2.23 \micron band, which reveals distinct surface heterogeneities essential for accurate feature matching.

Since the LEISA observations were acquired at varying spacecraft ranges, the native spatial resolution differs between files (see Table \ref{tbl2}). To facilitate pixel-by-pixel spectral analysis, it is essential to standardize the spatial grid. We therefore resampled all projection maps to a common resolution of 7 km/pixel. This ensures that a given pixel coordinate across all maps corresponds to the same geographic location on Pluto, thereby preserving the spectral integrity of individual geological features.

\subsection{Temperature Dependence of Optical Constants}
The \N{}-ice absorption feature at 2.148 \micron is highly sensitive to temperature and is commonly utilized as a spectral thermometer. Notably, prior to the New Horizons mission, this band was used to constrain Pluto's average surface temperature to approximately 40 K \citep{Tryka1994N2Pluto}.

We utilized laboratory refractive index data available at various temperatures (Table \ref{tbl1}). For \N{}-ice, two distinct datasets cover the relevant range: \citet{Grundy1993N2} (33--43 K) and \citet{Tryka1995N2Temp} (35.6--58.3 K). To model a broader thermal range, we concatenated the temperature coverage by selecting the data source based on the temperature regime: we used the spectra from \citet{Grundy1993N2} strictly for temperatures between 33 and 43 K, and the spectra from \citet{Tryka1995N2Temp} for temperatures between 43 and 58.3 K. Consequently, we constrained our thermal retrieval model to the range of 30--60 K.

It is important to note that \N{} undergoes a phase transition at 35.6 K, existing in the hexagonal $\beta$-phase above this temperature and the cubic $\alpha$-phase below it \citep{ProkhvatilovPhase}. Since our combined dataset primarily samples the $\beta$-phase, results in the 30--36 K range carry increased uncertainty. Furthermore, recent work by \citet{RAPOSA2026} suggests that the precise $\alpha-\beta$ transition boundary may deviate from the standard 35.6 K value depending on impurity content. 

For tholins (\TT{}), a single temperature-independent spectrum was employed. To estimate optical constants at intermediate temperatures, we applied cubic spline interpolation between the available laboratory measurements. For temperatures outside the measured range (specifically for \N{} and \CH{} dissolved in \N{}), we estimated the spectral behavior via linear extrapolation from the nearest available temperature data.

\subsection{Surface Parameter Retrieval}
For each surface element in the projected maps, we obtained up to five independent reflectance spectra (Fig. \ref{FIG:map_proj}C), allowing for a robust pixel-by-pixel analysis. Each spectral data point includes the independent associated measurement error (converted to reflectance units) and the photometric geometry (incidence, emission, and phase angles).

As noted in section 2.3, the original LEISA data contain two spectral segments with differing resolutions (Fig. \ref{FIG:map_proj}A). However, due to severe intensity artifacts (saturation and stray light) present in the high-resolution channel (2.1--2.25 \micron; see Fig. \ref{FIG:flat}), we excluded this segment from our analysis. Consequently, a surface element fully scanned by the LEISA instrument yields 198 spectral measurements per observation. For regions covered by all five FITS files, this results in a total of 990 data points per pixel. Given the selected map resolution of 7 km/pixel, the input dataset comprises five 3D arrays with dimensions $1067 \times 534 \times 990$ (encoding reflectance, error, incidence, emission, and phase).

To retrieve the surface properties, we applied the spectral model described in section~2.2, which simulates radiance factor (reflectance) as a function of material fractional area, effective particle diameter, temperature, and geometry. We developed a C++ implementation of the Levenberg-Marquardt algorithm to minimize the residual between the model and observations. The retrieval solves for 10 free parameters per pixel:

\begin{itemize}
    \item $F_i$ (4 parameters): Fractional areal coverage for each component (\CH{}-rich ice, \N{}-rich ice, \HO{}-ice, and \TT{}).
    \item $D_i$ (4 parameters): Effective particle diameter for each component.
    \item $f_{\chf:\nf}$ (1 parameter): The concentration of \CH{} dissolved in \N{} ice (constrained to $< 5\%$).
    \item $T$ (1 parameter): Surface temperature (constrained to the range 30--60 K).
\end{itemize}
We restricted our analysis to surface pixels with a minimum of 12 valid spectral measurements; typically, the data coverage ranged from 198 to 990 points per pixel. To mitigate errors arising from low signal-to-noise ratios and topographic shadowing, we excluded measurements acquired at incidence or emission angles exceeding $85^{\circ}$. 

We applied a weighting function to the residuals, defined as:
\begin{equation}
\label{eq.w}
W(x,y,\lambda)=W_\lambda(\lambda)\:\:cos\left(i(x,y,\lambda)\right)\:\:cos\left(e(x,y,\lambda)\right)
\end{equation}
where $i$ and $e$ represent the incidence and emission angles, respectively. To enhance the model's sensitivity to the nitrogen absorption feature, the spectral weight $W_\lambda(\lambda)$ was set to 1.5 for the range $\lambda = 2.11\text{--}2.18\,\mu\text{m}$, and 1.0 for all other wavelengths.

We employed the Levenberg-Marquardt algorithm to minimize two distinct objective functions: the sum of squared residuals (least squared $LS$, Eq. \ref{eq.ls}) and the reduced $\chi^2$ characteristic (Eq. \ref{eq.x2}):

\begin{equation}
\label{eq.ls}
LS=\frac{1}{N-10}\sum_\lambda W(\lambda)\left[\frac{I}{F}(\lambda,\mathbf{P})-\left(\frac{I}{F}\right)_\lambda\right]^2
\end{equation}

\begin{equation}
\label{eq.x2}
\chi^2=\frac{1}{N-10}\sum_\lambda W(\lambda)\left[\frac{\frac{I}{F}(\lambda,\mathbf{P})-\left(\frac{I}{F}\right)_\lambda}{\delta\left(\frac{I}{F}\right)_\lambda}\right]^2
\end{equation}

Here, $\frac{I}{F}(\lambda,\mathbf{P})$ denotes the modeled radiance factor as a function of the parameter vector $\mathbf{P}=\{F_{1-4},D_{1-4},f_{\chf:\nf},T\}$; $\left(\frac{I}{F}\right)_\lambda$ is the observed measurement at wavelength $\lambda$; $\delta\left(\frac{I}{F}\right)_\lambda$ is the measurement error; and $N$ is the total number of measurements (with $N-10$ representing the degrees of freedom).

While $\chi^2$ minimization is the standard approach in spectral fitting, we found that minimizing the  least squares (Eq. \ref{eq.ls}) yielded physically more plausible results in certain regions. Specifically, the $\chi^2$ method produced a sharp, likely artificial discontinuity in \CH{} abundance at the north pole (see Fig. \ref{FIG:x2}). In contrast, the least squares method resulted in a smoother and more continuous parameter distribution.

\section{Results and Discussion}

\begin{figure*}
	\centering
	\includegraphics[width=1.0\linewidth]{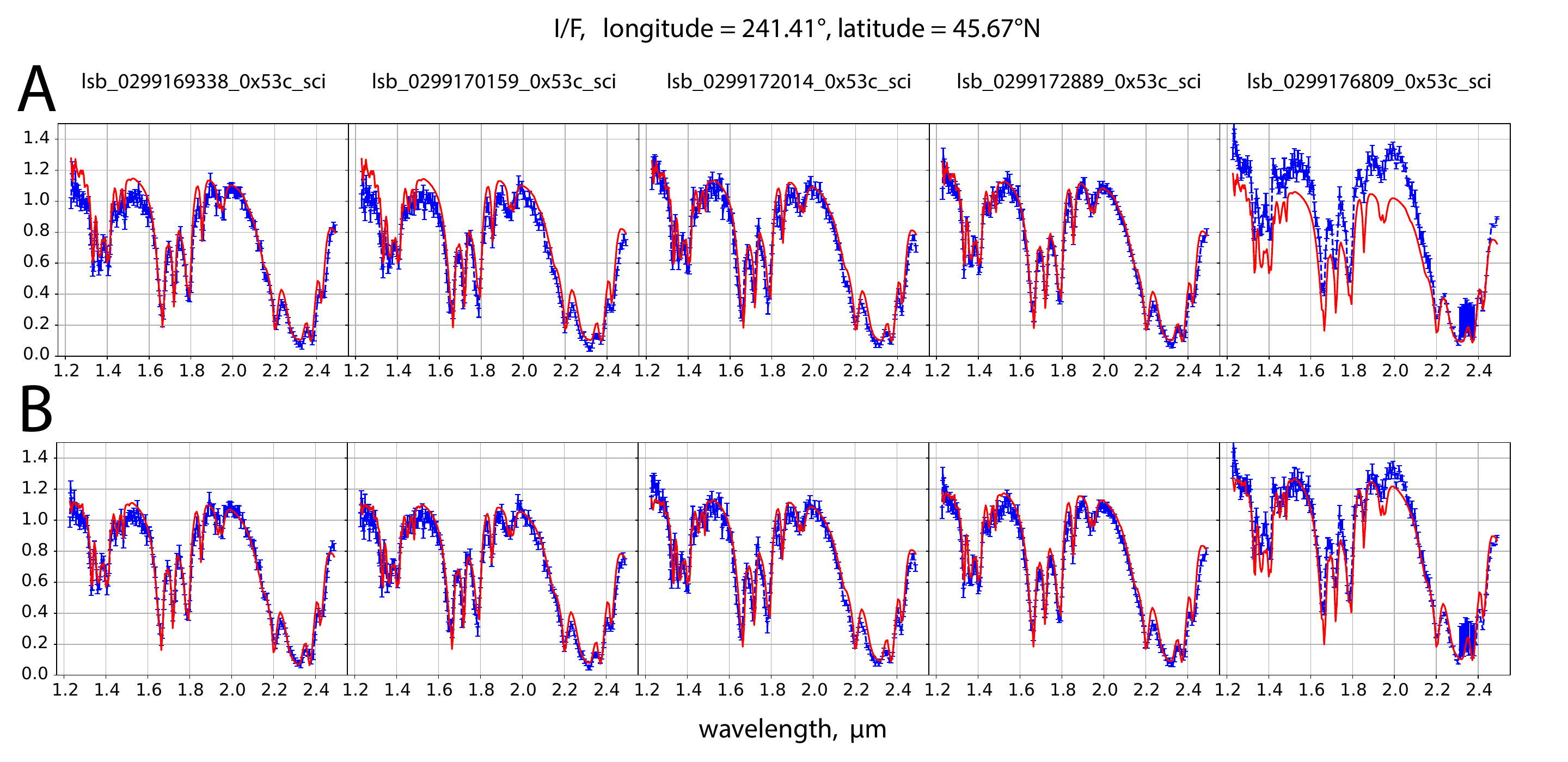}
	\caption{Illustration of scan-to-scan spectral level mismatch and the role of per-observation scale/offset terms. Blue: measured reflectance spectrum for an example surface element; red: best-fit model spectrum. (A) Fit without allowing per-scan scale and offset parameters ($S_j,O_j$), showing a systematic overestimation of the measured level for this scan. (B) Fit after solving for ($S_j,O_j$), which removes the bulk of the level mismatch while preserving absorption-band structure. Similar level offsets are observed across many pixels in the closest scan, motivating the inclusion of ($S_j,O_j$) as nuisance parameters to isolate compositional/temperature information from calibration artifacts.}
	\label{FIG:spectrum}
\end{figure*}

\begin{figure}
	\centering
	\includegraphics[width=0.7\linewidth]{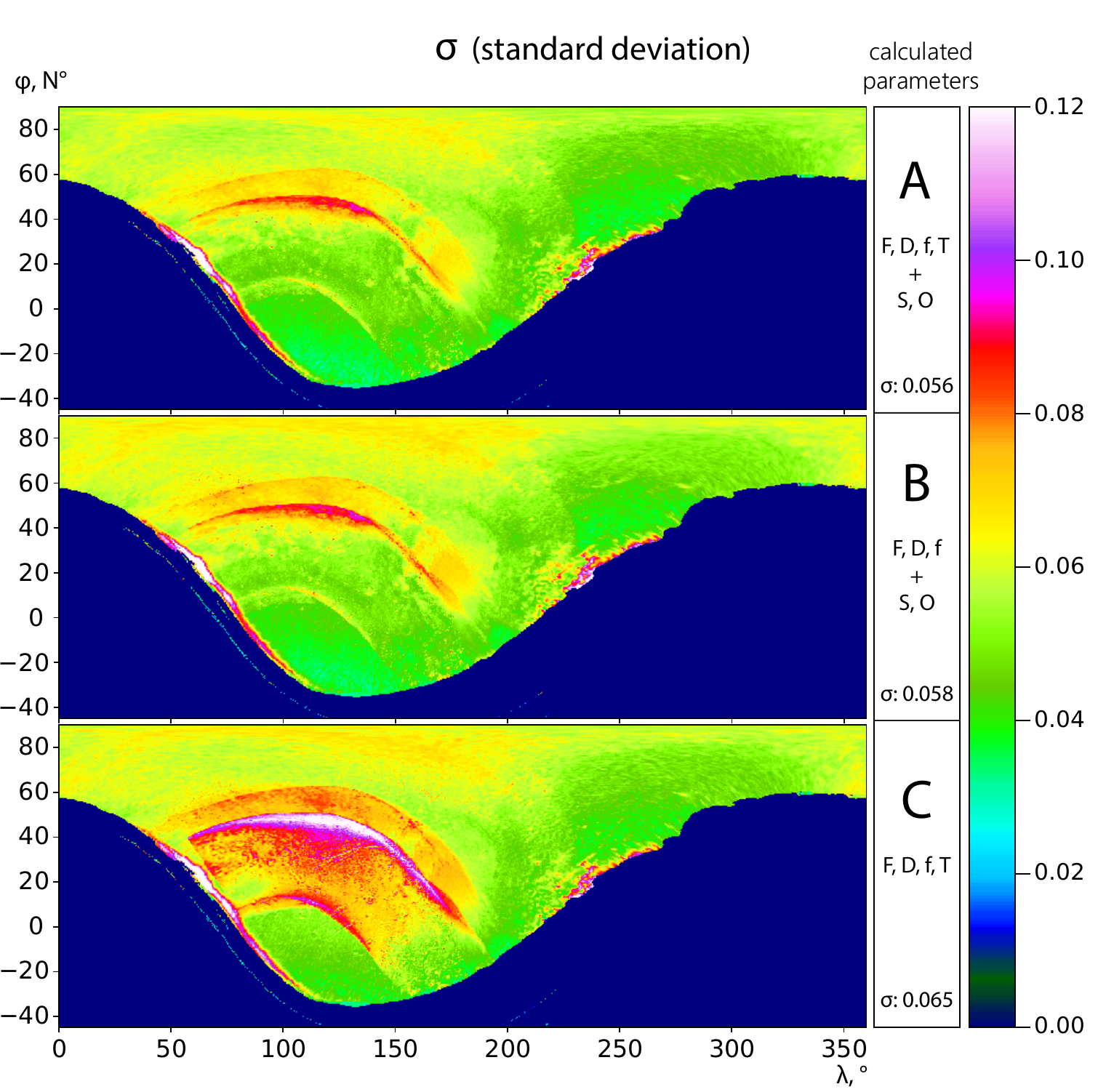}
	\caption{Goodness-of-fit (RMS) maps highlighting where additional parameters improve the spectral FITS and where residual artifacts remain. (A) RMS map for the full model, including fractional area, effective particle diameter, temperature, and per-scan scale/offset parameters ($F_i,D_i,f,T,S_j,O_j$). (B) Same as (A) but with temperature fixed (40K, no $T$ retrieval), showing a marked RMS increase (3.5\%) in the methane-rich northern terrains, consistent with temperature-sensitive band shapes contributing to the fit quality there. (C) Same as (A) but without allowing per-scan scale/offset ($S_j,O_j$), demonstrating that scan-to-scan level mismatches dominate the RMS in many regions and especially near the edges of the closest scan (\texttt{lsb\_0299176809\_0x53c\_sci}) and in scan overlap boundaries. $\phi$ and $\lambda$ are North latitude and East longitude, respectively.}
	\label{FIG:rms}
\end{figure}

Preliminary calculations revealed significant discrepancies in the measured reflectance for identical surface elements across different observations (Fig. \ref{FIG:spectrum}A). Notably, the spectral continuum level in the high-resolution, short-range observations is systematically higher—by up to 20\%—than in the long-range data. The largest deviation between the model and observations occurs in the file acquired at the closest approach (\texttt{lsb\_0299176809\_0x53c\_sci}), where radiance factor frequently exceed physical limits ($I/F > 1.3$). These artifacts are clearly visible in the root-mean-square (RMS) error map (Fig. \ref{FIG:rms}C).

Applying the original flat-field exacerbated the model-data mismatch, increasing the global RMS by approximately 10\%. We attempted to mitigate this by refining the macroscopic Hapke parameters ($g, h, B_0, \overline{\theta}$). However, even when fitting these parameters independently for each material (introducing $4\ \text{par.}\times4\ \text{mat.}=16$ additional parameters), the RMS residuals did not improve significantly. While the Hapke model can theoretically account for $I/F > 1$ via the opposition effect and anisotropic phase function, the observed inconsistencies persisted. Consequently, we attributed these discrepancies to residual radiometric calibration errors in the FITS files. To correct for this, we introduced empirical scale and offset factors for each observation:

\begin{equation}
\left(\frac{I}{F}\right)_{\text{corr}}(x,y,\lambda) = S_j \cdot \frac{I}{F}(x,y,\lambda) + O_j,
\end{equation}
where $S_j$ and $O_j$ represent the multiplicative scale and additive offset parameters for the $j$-th FITS file, respectively.

This approach introduces 10 global parameters (2 for each of the 5 files), which we optimized using the Levenberg-Marquardt algorithm to minimize the total RMS across all pixels. The final optimization procedure employed an alternating iterative scheme:
\begin{enumerate}
    \item Refine surface parameters ($F_i, D_i, f, T$) for each pixel while holding calibration factors constant.
    \item Refine calibration factors ($S_j, O_j$) globally while holding surface parameters constant.
    \item Repeat until convergence.
\end{enumerate}

All computations were performed on a workstation equipped with a 4-core AMD Ryzen 5 3550H processor and 16 GB of RAM. The code was implemented in C++ using Microsoft Visual Studio Community 2022 and compiled with the Intel C++ 2025 compiler (Intel oneAPI Base Toolkit). A single iteration of the surface parameter retrieval ($F_i, D_i, f, T$) for the entire dataset requires approximately 60 minutes with 4-core parallelization, while the calibration update ($S_j, O_j$) takes approximately 2 minutes.

As discussed in section~2.6, we performed parallel retrievals minimizing both the least squares (Eq. \ref{eq.ls}) and $\chi^2$ (Eq. \ref{eq.x2}) objective functions, applying the weighting function (Eq. \ref{eq.w}) in both cases. The results diverge most significantly in the northern polar region (Fig. \ref{FIG:x2}). The $\chi^2$ minimization produces a sharper contrast in composition, characterized by a higher concentration of \CH{} and a corresponding depletion of \N{}. Specifically, in the $60^\circ\text{N}$--$90^\circ\text{N}$ latitude band, the mean abundances derived via $\chi^2$ minimization are $76\%$ for \CH{} and $14\%$ for \N{}. In contrast, the least squares approach yields a smoother gradient with mean values of $69\%$ and $20\%$, respectively. This difference in composition propagates to the thermal retrieval: the global average surface temperature is $53$ K for the $\chi^2$ solution, compared to $50$ K for the least squares solution due to the higher concentration for \N{} in polar region.

The subsequent analysis focuses on the results obtained via least squares minimization (Eq. \ref{eq.ls}). The derived distributions of ice composition and effective particle diameter (Fig. \ref{FIG:F}, Fig. \ref{FIG:D}) are consistent with the findings reported by \citet{Protopapa2017}. However, residual artifacts arising from the mosaicking of overlapping FITS observations remain visible, particularly in the \TT{} distribution map.

The derived scale and offset parameters ($S_j, O_j$) are listed in Table \ref{tbl_so}. We recognize that the use of linear scaling factors is a heuristic approach; however, it enables the simultaneous analysis of the complete dataset. Notably, the significant scale factor (1.215) required for the \texttt{lsb\_0299176809\_0x53c\_sci} observation indicates potential inaccuracies in its radiometric calibration, as also illustrated in Figure \ref{FIG:spectrum}.

\begin{table}
\caption{Per-scan scale/offset parameters, calculated by least square minimization (Eq.~\ref{eq.ls}).}
%\centering
\footnotesize
\begin{tabular}{l l p{2.0cm}}
\hline
File name & $S_j$ & $O_j$ \\
\hline
\\
lsb\_0299169338\_0x53c\_sci & 1.009 & -2.465e-02 \\
\\
lsb\_0299170159\_0x53c\_sci & 0.989 & -1.322e-02 \\
\\
lsb\_0299172014\_0x53c\_sci & 1.015 &  1.763e-04 \\
\\
lsb\_0299172889\_0x53c\_sci & 1.067 & -6.867e-03 \\
\\
lsb\_0299176809\_0x53c\_sci & 1.215 &  2.230e-03 \\
\\
\hline
\end{tabular}
\label{tbl_so}
\end{table}

The retrieved temperature distribution (Fig. \ref{FIG:T}) presents a complex picture. In \N{}-rich regions, the temperature is robustly constrained (averaging $40$ K) due to the high thermal sensitivity of the $2.148\,\mu$m absorption band. Conversely, in regions dominated by tholins (\TT{}), the retrieved temperature is unreliable due to the lack of temperature-dependent laboratory spectra for these materials.

Of particular interest is the northern polar region, which is characterized by elevated \CH{} abundance. Here, the retrieved temperature frequently hits the model's upper bound of $60$ K. Although treating temperature as a free parameter significantly reduced the RMS residuals in this region (by up to $3.5\%$; see Fig. \ref{FIG:rms}A vs. Fig. \ref{FIG:rms}B), this result is likely a model artifact rather than a physical reality. Absorption bands typically broaden with increasing temperature; however, similar broadening effects can be induced by physical factors not included in our model, such as wide grain-size distributions, radiation-induced crystal damage, or lattice impurities. Consequently, the optimization algorithm may artificially drives the temperature to the upper limit to mimic band broadening caused by these unmodeled effects.

We also observed a strong sensitivity of the retrieved temperature and fractional area of \N{} abundance within Sputnik Planitia to the assumed imaginary refractive index ($k$) in the 1.2--2.1 \micron \N{} transparency window. Our baseline model assumes $k=0$ for \N{} at wavelengths $\lambda < 2.1\,\mu$m. However, as shown in Fig. \ref{FIG:k}, introducing small non-zero values for $k$ in this continuum region significantly alters the retrieved material distribution in Sputnik Planitia. While the assumption of negligible absorption is physically motivated, laboratory measurements by \citet{Grundy1993N2} and \citet{Tryka1995N2Temp} indicate that the wing of the 2.148 \micron absorption band does not asymptotically approach zero, maintaining a small residual value. Furthermore, absorption in this transparency window may also be driven by carbon monoxide (CO), which is known to be present in Sputnik Planitia \citep{GrundyCO} but was not explicitly included in this model.

\section{Conclusions}

We produced co-registered, 7 km/pixel equirectangular maps of Pluto's surface composition and temperature by combining five close-approach LEISA observations and performing pixel-by-pixel spectral inversion in a Hapke-based mixing framework. The resulting distributions of \CH{}-rich ice, \N{}-rich ice, \HO{} ice, and \TT{} are broadly consistent with previous LEISA mapping efforts, while adding a simultaneous temperature retrieval constrained primarily by the \N{} absorption near 2.148~\micron.

A key outcome of this work is that the remaining fit residuals cannot be attributed to a single cause. The Hapke + equivalent-slab formulation, while physically motivated and capable of reproducing the main absorption-band structure, does not fully reconcile the spectral differences observed for the same surface elements across multiple scans and viewing geometries. This limitation likely reflects a combination of model incompleteness (e.g., unmodelled grain-size distributions, missing temperature dependence for tholin optical constants, and neglected absorbers such as CO in \N{}-rich terrains) and residual observation-level calibration uncertainties.

In particular, the closest-approach LEISA scan \texttt{lsb\_0299176809\_0x53c\_sci} shows clear signatures of radiometric inconsistency relative to the other observations: measured reflectance values frequently exceed physically plausible levels (often reaching $I/F \ge 1.3$ in our products), and the global per-scan scaling required to bring this scan into agreement with the rest of the dataset is the largest among the five ($S_j=1.215$; Table~\ref{tbl_so}). Together, these indicators suggest that this scan suffers from residual calibration issues beyond what can be absorbed by physically meaningful surface parameters alone.

The preprocessing and co-registration pipeline developed here (bad-pixel handling, offset subtraction, empirical flat-field refinement, and map matching) substantially reduces edge artifacts and discontinuities in overlap regions and enables joint analysis of the full dataset. However, the need for per-scan scale/offset terms and the behavior of \texttt{lsb\_0299176809\_0x53c\_sci} emphasize that further improvements in radiometric calibration and/or a more flexible spectral-physical model will be required to fully exploit multi-scan LEISA mosaics for quantitative thermophysical and compositional interpretation.

Uncertainty estimates for the retrieved parameters were computed and are provided with the released map products. We note that these uncertainties are limited by the Levenberg--Marquardt framework and can be imperfect in some cases, particularly where numerical derivatives are used or the model is close to ill-conditioned.

\section{Data Availability}

The derived parameter maps (Figs. \ref{FIG:F}, \ref{FIG:D}, \ref{FIG:T}), and their errors, the 7 km/pixel equirectangular projections for the five LEISA observations, and the radiometrically corrected FITS files are publicly available via Zenodo (\url{https://doi.org/10.5281/zenodo.18825240}). To facilitate the assessment of residual calibration artifacts, the repository also includes the empirical flat-field correction maps ($f^*(x,y)$) and the per-observation radiometric scaling factors ($S_j, O_j$).

\section{Acknowledgments}
We would like to express our special gratitude to A. A. Lomakin from the Space Research Institute of the Russian Academy of Sciences for his expert consultation.

This work was supported by M.V.Lomonosov
Moscow State University Program of Development.

\printcredits

\bibliographystyle{cas-model2-names}
\bibliography{cas-refs-r2}

\appendix
\section{Appendix}

\begin{figure*}
	\centering
	\includegraphics[width=1.0\linewidth]{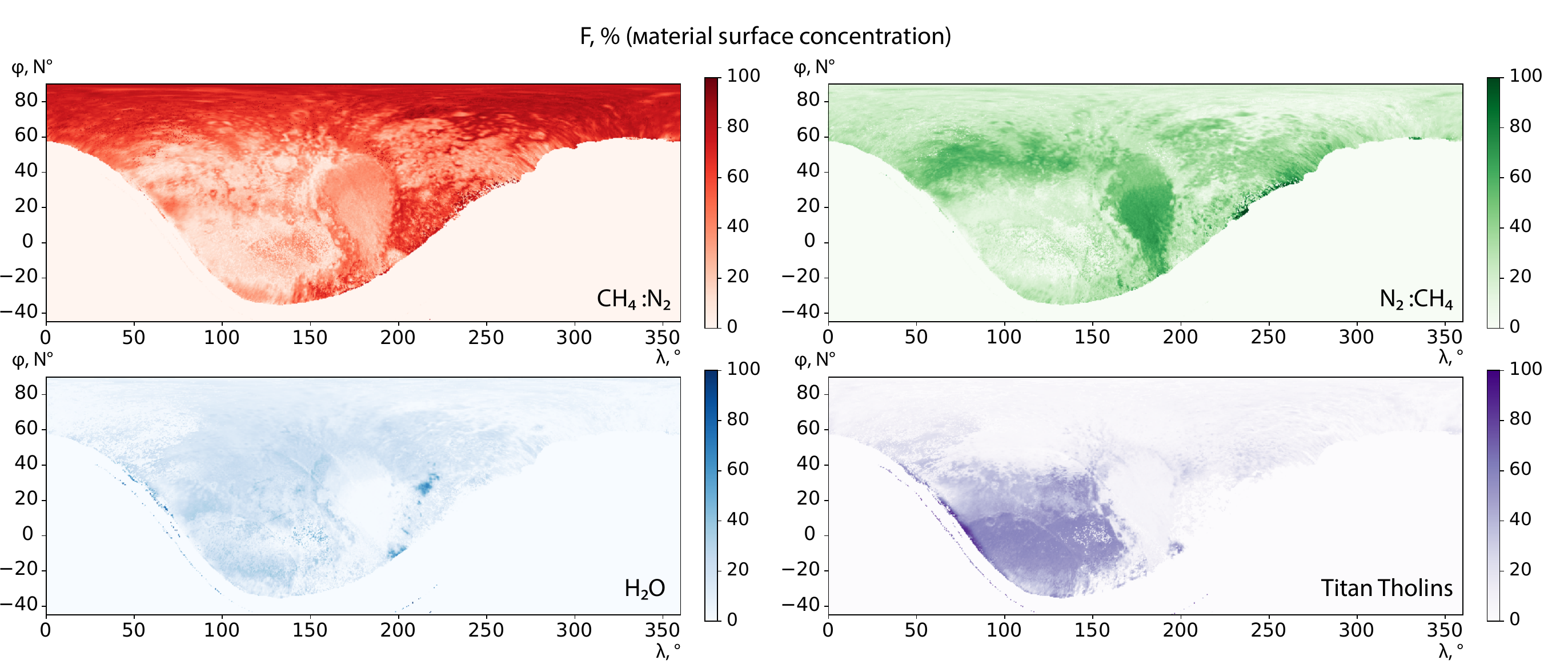}
	\caption{Maps of materials fractional area ($F_i$) in \% for \CH{}, \N{}, \HO{}, \TT{}. $\phi$ and $\lambda$ are North latitude and East longitude, respectively.}
	\label{FIG:F}
\end{figure*}
\begin{figure*}
	\centering
	\includegraphics[width=1.0\linewidth]{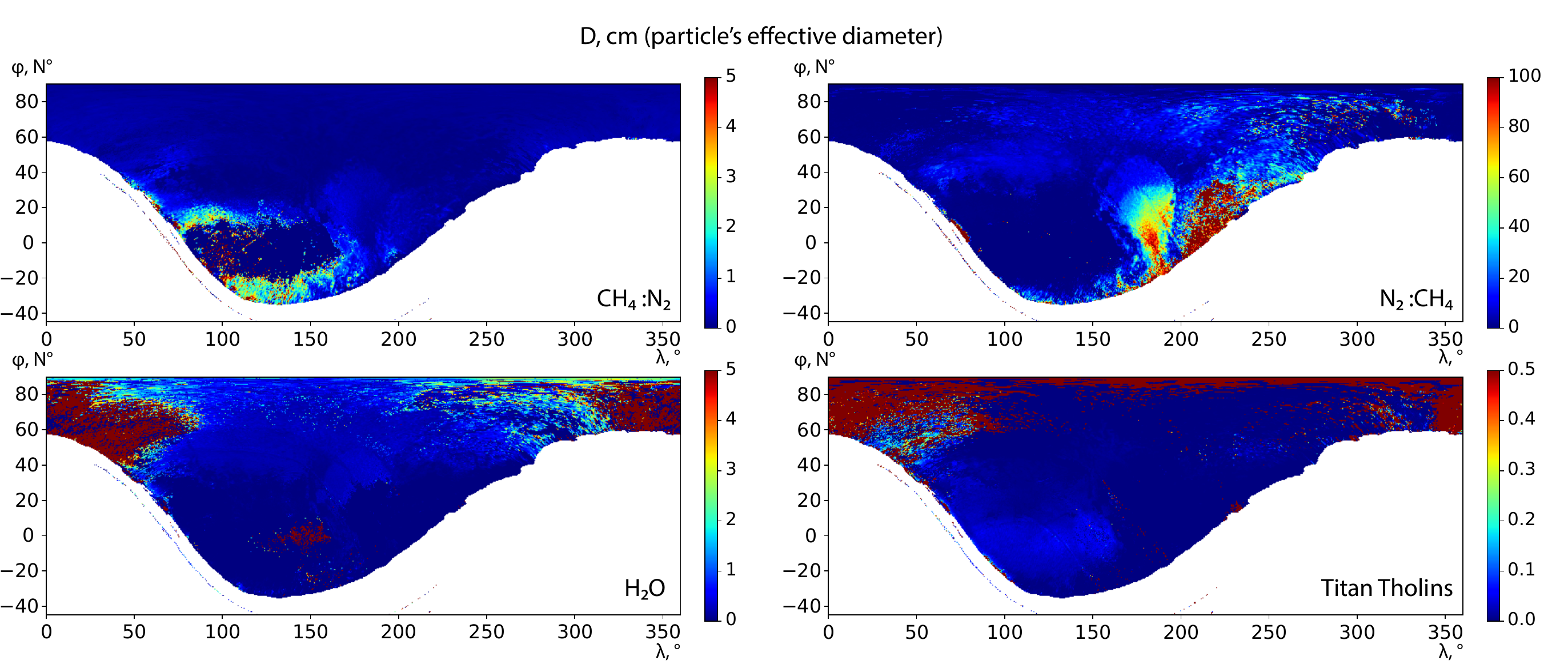}
	\caption{Maps of effective particle diameters ($D_i$) for \CH{}, \N{}, \HO{}, \TT{}. The following limits for diameters were used in the calculations: \CH{} - 30 cm, \N{} - 100 cm, \HO{} - 20 cm, \TT{} - 1 cm. $\phi$ and $\lambda$ are North latitude and East longitude, respectively.}
	\label{FIG:D}
\end{figure*}
\begin{figure*}
	\centering
	\includegraphics[width=1.0\linewidth]{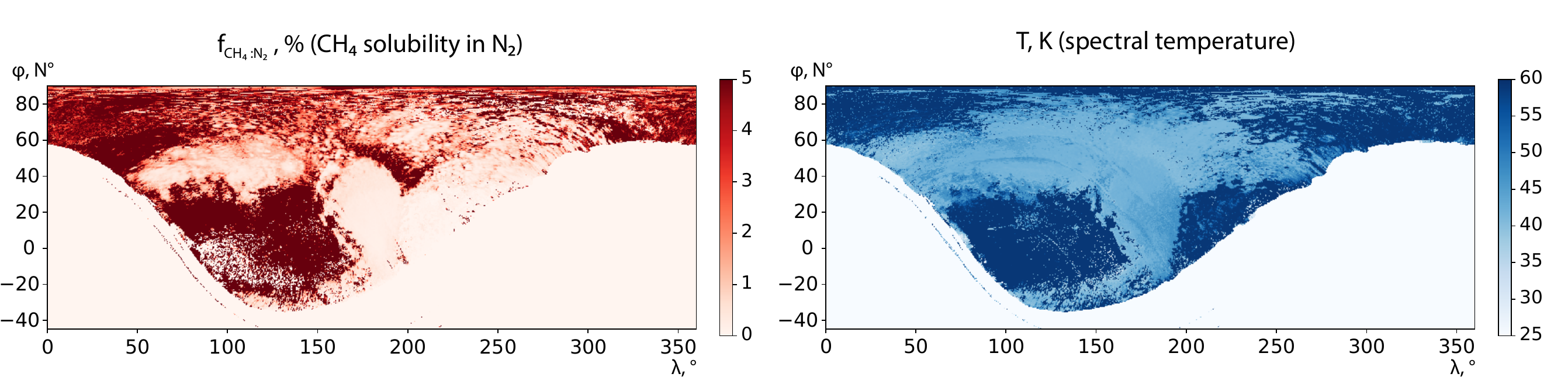}
	\caption{Solubility map of \CH{} in \N{} solution ($f_{\chf:\nf}$) and temperature distribution map ($T$). $\phi$ and $\lambda$ are North latitude and East longitude, respectively.}
	\label{FIG:T}
\end{figure*}

\begin{figure*}
	\centering
	\includegraphics[width=1.0\linewidth]{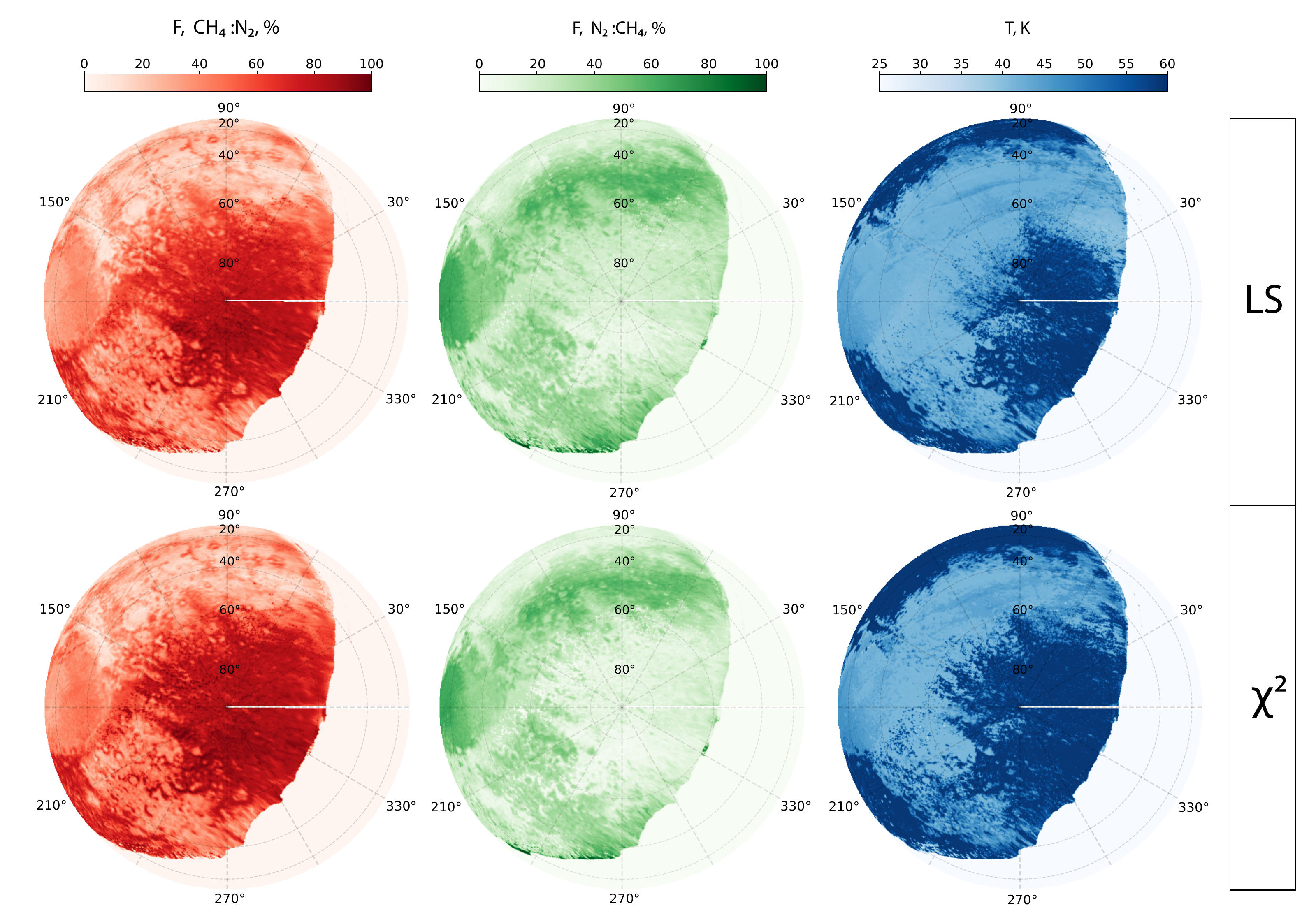}
	\caption{Comparison of Northern Hemisphere maps showing \CH{}, \N{}, and temperature distributions derived using least squares minimization (LS, Eq.~\ref{eq.ls}) and $\chi^2$ minimization (Eq.~\ref{eq.x2}). With the $\chi^2$ method, the mean abundances of \CH{} and \N{} in the $60^\circ\text{N}$--$90^\circ\text{N}$ region are $76\%$ and $14\%$, respectively, with an average surface temperature of 53\,K. The LS minimization yields values of $69\%$ for \CH{} and $20\%$ for \N{}, with an average temperature of 50\,K. The numerical values on the projections correspond to North latitude and East longitude.}
\label{FIG:x2}
\end{figure*}

\begin{figure*}
	\centering
	\includegraphics[width=1.0\linewidth]{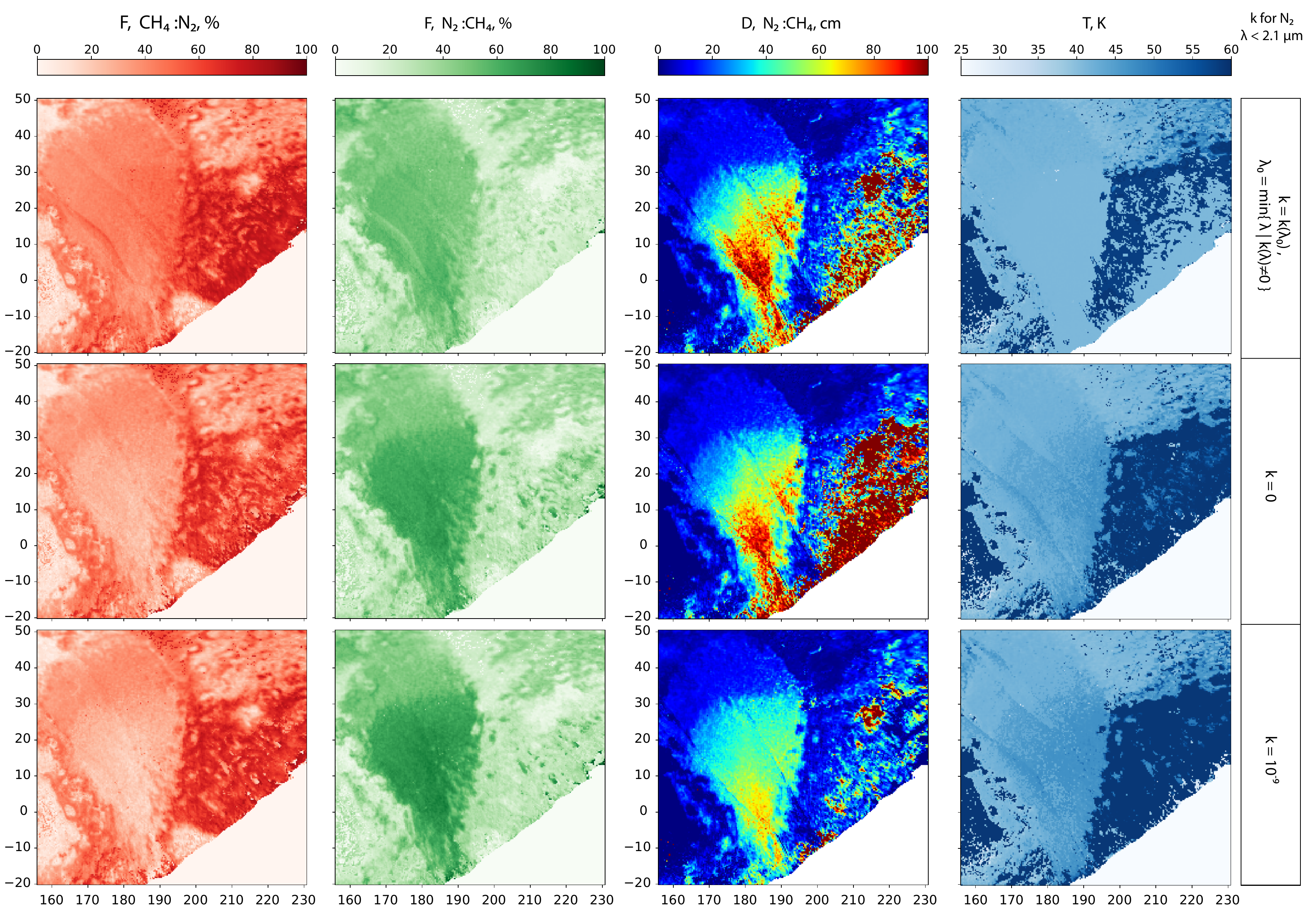}
	\caption{Retrieval results for \CH{}, \N{}, \N{} grain size, and temperature for different assumptions of the \N{} refractive index imaginary part ($k$) in the transparency window ($\lambda < 2.1$\,\micron). The Sputnik Planitia region is shown due to its high nitrogen ice abundance. The rows correspond to different $k$ values: (top) $k$ is fixed to the last non-zero value of the \N{} absorption band tail; (middle) $k=0$; and (bottom) $k=10^{-9}$ for all temperatures. The numerical coordinates on the projections correspond to North latitude and East longitude.}
	\label{FIG:k}
\end{figure*}

\end{document}